\documentclass[11pt]{article}

\usepackage{float}
\usepackage[T1]{fontenc}
\usepackage[utf8]{inputenc}
\usepackage{lmodern}
\usepackage{microtype}
\usepackage[linewidth=1pt]{mdframed}
\usepackage{url}

\usepackage[bookmarks]{hyperref}
\usepackage{pdfpages}
\usepackage{epsfig}
\usepackage{url}
\usepackage{graphicx,color}% Include figure files
\usepackage[psamsfonts]{amssymb}
\usepackage{amsmath}
\usepackage{indentfirst}
\usepackage{mathrsfs}
\usepackage{booktabs}
\usepackage{tabularx}
\usepackage{setspace}
\onehalfspace
\usepackage{color}

\newcommand{\tableCaptionNegativeVSpace}{\vspace{-1em}}

\title{Are Bitcoin Bubbles Predictable?\\ Combining a Generalized Metcalfe's Law and the LPPLS Model}
\author{Spencer Wheatley$^{1*}$, Didier Sornette$^{1,2*}$, Tobias Huber$^{1}$, Max Reppen$^{3}$, and Robert N. Gantner\\
$^1$ {\small ETH Zurich, Department of Management, Technology and Economics, Switzerland}\\
$^2$ {\small Swiss Finance Institute, c/o University of Geneva, Switzerland}\\
$^3$ {\small ETH Zurich, Department of Mathematics}\\
{\small e-mails: swheatley@ethz.ch, dsornette@ethz.ch}\\
$^*${\small corresponding authors}}

\begin{document}
\maketitle

\begin{abstract}
We develop a strong diagnostic for bubbles and crashes in bitcoin, by analyzing the coincidence (and its absence) of fundamental and technical indicators. Using a generalized Metcalfe's law based on network properties, a fundamental value is quantified and shown to be heavily exceeded, on at least four occasions, by bubbles that grow and burst. In these bubbles, we detect a universal super-exponential unsustainable growth.
We model this universal pattern with the Log-Periodic Power Law Singularity (LPPLS) model, which parsimoniously captures diverse positive feedback phenomena, such as herding and imitation. The LPPLS model is shown to provide an ex-ante warning of market instabilities, quantifying a high crash hazard and probabilistic bracket of the crash time consistent with the actual corrections; although, as always, the precise time and trigger (which straw breaks the camel's back) being exogenous and unpredictable. Looking forward, our analysis identifies a substantial but not unprecedented overvaluation in the price of bitcoin, suggesting many months of volatile sideways bitcoin prices ahead (from the time of writing, March 2018).
\end{abstract}

\pagebreak
\clearpage

\section{ Introduction}

In 2008, pseudonymous Satoshi Nakamoto introduced the digital decentralized cryptocurrency, bitcoin~\cite{nakamoto2008bitcoin}, and the innovative blockchain technology that underlies its peer-to-peer global payment network\footnote{In this network, transactions, which do not rely on an intermediary, are verified by network nodes and, through cryptography, immutably recorded in a decentralized publicly distributed ledger~\cite{antonopoulos2014mastering}.}.
 Since its techno-libertarian beginnings, which envisioned bitcoin as an alternative to the central banking system, bitcoin has experienced super-exponential growth. Fueled by the rise of bitcoin, a myriad of other cryptocurrencies have erupted into the mainstream with a range of highly disruptive use-cases foreseen. Cryptocurrencies have become an emerging asset class~\cite{burniske2018cryptoassets}. At the end of 2017, the price of bitcoin peaked at almost 20'000 USD, and the combined market capitalization of cryptocurrencies reached around 800 billion USD. 

The explosive growth of bitcoin intensified debates about the cryptocurrency's intrinsic or fundamental value. While many pundits have claimed that bitcoin is a scam and its value will eventually fall to zero, others believe that further enormous growth and adoption await, often comparing to the market capitalization of monetary assets, or stores of value. By comparing bitcoin to gold, an analogy that is based on the digital scarcity that is built into the bitcoin protocol, some markets analysts predicted bitcoin prices as a high as 10 million per bitcoin~\cite{Lee2017}. Nobel laureate and bubble expert, Robert Shiller, epitomized this ambiguity of bitcoin price predictions when he stated, at the 2018 Davos World Economic Forum, that ``bitcoin could be here for 100 years but it's more likely to totally collapse'' and, ``you just put an upper bound on [bitcoin] with the value of the world's money supply. But that upper bound is awfully big. So it can be anywhere between zero and there.''~\cite{CNBC2018}.

There is an emerging academic literature on cryptocurrency valuations~\cite{Gertchev2013, harwick2016, bergstra2014, Yermack2015, jenssen2014, van2014,bouoiyour2015does, polasik2015price,zhang2014economics} and their growth mechanisms~\cite{wu2018classification}. Many of these studies attribute some technical feature of the bitcoin protocol, such as the ``proof-of-work'' system on which the bitcoin cryptocurrency is based, as a source of value\footnote{The question of what constitutes the value of money has preoccupied generations of thinkers. About 2050 years ago, Aristotle was probably the first to argue that money needed a high cost of production in order to make it valuable. In other words, according to Aristotle, the larger the effort to create new money, the more valuable it is. This was later elaborated into the labor theory of value, starting with Adam Smith, David Ricardo, and becoming the central thesis of Marxian economics. Nowadays, this concept is archaic and it is well understood that money is credit (see e.g.,~\cite{von2017should}). It is thus puzzling that cryptocurrencies with proof-of-work designs, which aim at revolutionizing money and exchanges between individuals, use a very old and obsolete concept that has been mostly abandoned in economics.}.
However, as has been proposed by former Wall Street analyst Tom Lee~\cite{Lee2017}, an early academic proposal  (see Ref.\cite{alabi2017digital}), by now widely discussed within cryptocurrency communities, is that an alternative valuation of bitcoin can be based on its network of users. In the 1980s, Metcalfe proposed that the value of a network is proportional to the square of the number of nodes~\cite{metcalfe2013metcalfe}. This may also be called the network effect, and has been found to hold for many networked systems. If Metcalfe's law holds here, fundamental valuation of bitcoin may in fact be far easier than valuation of equities \footnote{See however Cauwels and Sornette~\cite{CauwelsSor12,ForroCauwelsSor12}, who developed an original valuation method for social network firms based on the economic value of the demographics of users, and were able to predict ex-ante the performance of companies such as Facebook, Zynga and Groupon after their IPO's.}---which relies on various multiples, such as price-to-earnings, price-to-book, or price-to-cash-flow ratios---and will therefore admit an indication of bubbles.

Although it seems relatively obvious that bubbles exist within cryptocurrencies, it is not a straw man argument that, in finance and economics, financial bubbles are often excluded based on market efficiency rationalization\footnote{For instance, the Efficient Market Hypothesis (EMH) assumes that prices quasi-instantaneously reflects all available information. Thus, market crashes result from novel very negative information that gets incorporated into prices~\cite{fama1970efficient}.}, which assume an unpredictable market price, for instance following a kind of geometrical random walk (see e.g.,~\cite{malkiel1999random}). In sharp contrast, Didier Sornette and co-workers claim that bubbles exist and are ubiquitous. Moreover, they can be accurately described by a deterministic nonlinear trend called the Log-Periodic Power Law Singularity (LPPLS) model, potentially with highly persistent, but ultimately mean-reverting, errors. The LPPLS model combines two well documented empirical and phenomenological features of bubbles (see~\cite{sornette2015financial} for a recent review): 
\begin{enumerate}
\item the price exhibits a transient faster-than-exponential growth (i.e., where the growth rate itself is growing)---resulting from positive feedbacks like herding~\cite{sornette2003stock}---that is modeled by a hyperbolic power law with a singularity in finite time, i.e., endogenously approaching an infinite value and therefore necessitating a crash or correction before the singularity is reached;

\item it is also decorated with accelerating log-periodic volatility fluctuations, embodying spirals of competing expectations of higher returns (bullish) and an impending crash (bearish)~\cite{johansen1999predicting,sornette1998discrete}.   Such log-periodic fluctuations are ubiquitous in complex systems with a hierarchical structure and also appear spontaneously as a result of the interplay between (i) inertia, (ii) nonlinear positive and (iii) nonlinear negative feedback loops~\cite{ide2002oscillatory}.
\end{enumerate}

The model thus characterizes a process in which, as speculative frenzy intensifies, the bubble matures towards its endogenous critical point, and becomes increasingly unstable, such that any small disturbance can trigger a crash. This has been further formalized in the so-called JLS model where the
rate of return accelerates towards a singularity, compensated by the growing crash hazard rate \cite{johansen1999predicting,johansen2000crashes}, providing a generalized return-risk relationship. We emphasize that one should not focus on the instantaneous and rather unpredictable trigger itself, but monitor the increasingly unstable state of the bubbly market, and prepare for a correction.

Here, we combine---as a fundamental measure---a generalized Metcalfe's law and---as a technical measure---the LPPLS model, in order to diagnose bubbles in bitcoin. When both measures coincide, this provides a convincing indication of a bubble and impending correction. If, in hindsight, such signals are followed by a correction similar to that suggested, they provide compelling evidence that a bubble and crash did indeed take place. 

This paper is organized as follows. In the first part, we document a generalized Metcalfe's law describing the growth of the population of active bitcoin users. We show that the generalized Metcalfe's law provides a support level, and that the ratio of market capitalization to ``the Metcalfe value'' gives a relative valuation ratio. On this basis, we identify a current substantial but not unprecedented overvaluation in the price of bitcoin. In the second part of the paper, we unearth a universal super-exponential bubble signature in four bitcoin bubbles, which corresponds to the LPPLS model with a reasonable range of parameters. The LPPLS model is shown to provide advance warning, in particular with confidence intervals for the critical bursting time based on profile likelihood. An LPPLS fitting algorithm is presented, allowing for selection of the bubble start time, and offering an interval for the crash time, in a probabilistically sound way. We conclude the paper with a brief discussion.

\section{Fundamental value of bitcoin: active users \& a generalized Metcalfe's law}

Metcalfe's law states that the value, in this case market capitalization (cap), of a network is,
\begin{equation}
\label{eqn:Metcalfslaw}
p= e^{\alpha_0} u^{\beta_0},\quad \beta_0=2,
\end{equation}
where $u$ is called the number of active users, imperfectly quantified by a proxy, being the number of active addresses\footnote{The data is collected from bitinfocharts.com. Limitations: It is difficult to know the true number of active users, in particular because a single user can have multiple addresses that, to an outsider, cannot be distinguished from addresses belonging to multiple users. Moreover, bitcoin.org's Developer Guide \cite{btc_dev_guide} discourages key reuse, advising that each key should only be used for two transactions (to receive, then send), and that all change should be sent to a new address, generated at the time of transaction (belonging to the sender). Depending on to what extent this advice is followed, this measure is thus an unclear mix between the number of \emph{daily} users and the number of \emph{daily} transactions (their activity).}. It is a single factor model for a fundamental valuation of bitcoin, and plausibly for other cryptocurrencies. From Figure 1, we indeed see a surprisingly clear log-linear relationship. Rather than taking Metcalfe's law as a given, we estimate the relevant parameters by a log-linear regression  model, which we refer to as the (generalized) Metcalfe law, 
\begin{equation}
\label{eqn:generalizedMetcalfsLaw}
\ln(p)=\alpha + \beta \ln(u) + \epsilon.
\end{equation}

The result of this fit, on 2'782 daily values, from 17-07-2010 to 26-02-2018, is a slope $\beta=1.69$ (standard error 0.0076), intercept $\alpha=1.51$ (0.087), and coefficient of determination $R^2=0.95$\footnote{Such high values are of limited value as one often obtains high coefficients of determination when regressing unrelated trending/non-stationary series onto each-other (so-called ``spurious regression''). In this case, the causal link between active users and market cap is assumed.}. Forcing the exponent $\beta$ to be equal to 2 would result in an intercept of $-2.01~(0.018)$, but this regression is significantly worse than the above\footnote{An ANOVA/F-test comparing the two models gives a p-value of less than $10^{-16}$. Further, the calibrated value of the slope, $\beta=1.69$, with standard error 0.0076, is clearly far from Metcalfe's value $2$. }. Further, a slope of 2 (or larger) is robustly rejected on moving windows\footnote{On 83\% of 1-year windows, the parameter $\beta$ is less than 2, and on 75\% of windows the parameter $\beta$ is significantly less than 2, at level $p=0.05$.}.  On this basis, it seems that the value 2 proposed by Metcalfe is too large, at least for the bitcoin ecosystem.\footnote{Note, however, that the measure of $u$ is overestimating the true number of {\emph{daily}} users. It is possible that this does affect the precise value of the exponent $\beta$. On the other hand, it could provide an underestimate of the number of active users if the typical user does not transact daily.}

\begin{figure}[h!]
\centerline{\includegraphics[width=18cm]{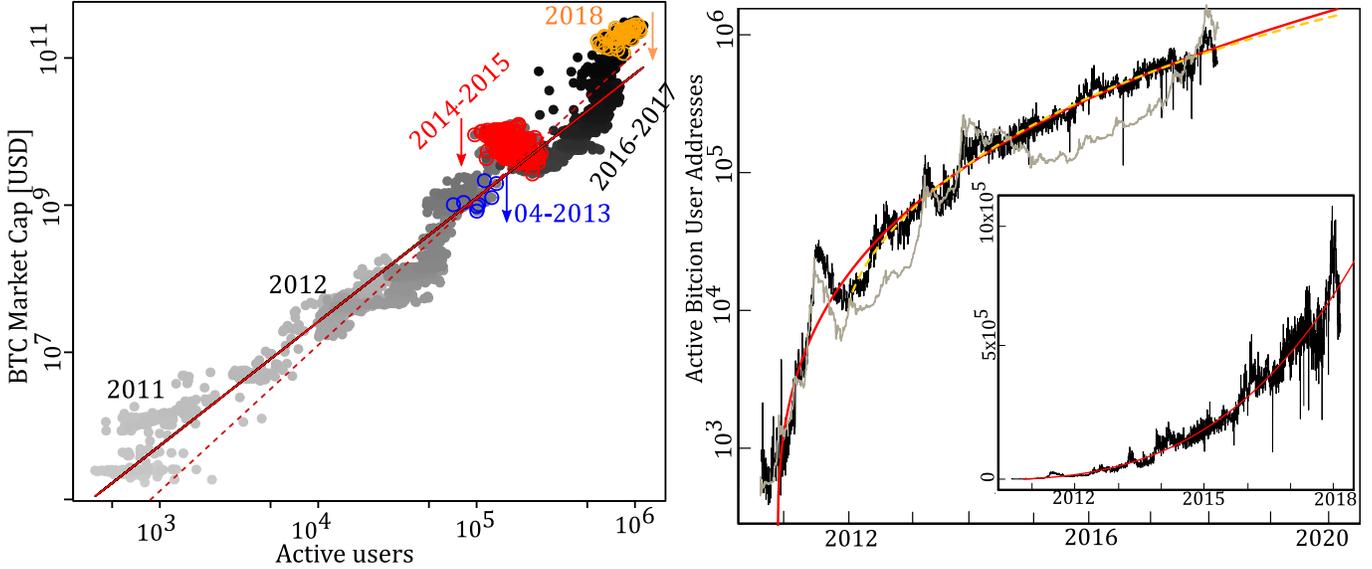}}
\caption{Left panel: Scatterplot of the bitcoin market cap versus the number of active users, with logarithmic scales. The points becomes darker as time progresses, and the three latest crashes are indicated by colored points, and arrows indicating the size of the correction. The generalized Metcalfe regression is given in solid red, and with slope forced to be 2 given by the dashed red line. Right panel: Active users (rough black line), again in a logarithmic scale, as a function of time, with linear scale inset. A scaled bitcoin market cap is overlaid with the grey line. The red and dashed yellow lines are the nonlinear regression fits of active users, fitting on different time windows.}
\label{fig:fig1}
\end{figure}

It should be noted that this regression severely violates the assumption that the errors be independent and identically distributed, as there are persistent deviations from the regression line. This statement deserves to be made in more salient terms: the residuals are in fact the bubbles and crashes! This is the focus of the second part of this paper. Ignoring this egregious violation of the so-called Gauss--Markov conditions is well known to give the false impression of precise parameter estimates. Further, endogeneity is an issue, as the number of active users may determine market cap in the long term, but large fluctuations in market cap can also plausibly trigger fluctuations in active users on shorter time scales (see Figure~1). We address this by smoothing active users\footnote{This is done with the R library \texttt{loess} with 5 equivalent degrees of freedom.}, assuming that this will average out the effects of short term feedback of market cap onto active users. A multiplier effect is also a plausible consequence of this endogeneity: a jump in user activity causes an increment in market cap, which triggers a (smaller) jump in user activity, feeding back into market cap, etc. Therefore, we do not claim to isolate the effect of a single increment in active users on market cap, and do not need it. Finally, we omit formal tests for causality, given the plausibility of the general mechanism behind Metcalfe's law, as well as the very turbulent and only long-term adherence to it\footnote{The exponent value 2 in the standard Metcalfe's law embodies the idea that the value of the network is proportional to the total number of interactions or exchanges, which themselves scales as the total number of possible connections. In other words, Metcalfe's law assumes full connectivity between all users. This does not seem realistic. Our finding of a smaller exponent $\beta \approx 1+2/3$ expresses a more sparsely connected network in which each user is on average linked to $\sim N^{2/3}$ other users in the total network of N users. For instance, for N=1 Million, a typical user is then connected to ``only'' 10'000 other users, a more realistic figure.}. 

In view of these limitations, the generalized Metcalfe's law here is still rather impressive, and will be shown to be highly useful, despite its radical simplicity and uncertain parameter values. Of course, one may add other variables to the regression, which further characterize the network, such as degree of centralization, transaction costs, volume, etc. However, the actual volume (value of authentic transactions) for instance is not only difficult to know, but, in general financial markets, is known to be highly correlated with volatility, of which bubbles and bursts are the most formidable contributors, and may therefore be too endogenous to soundly indicate a fundamental value.
Therefore, the variable `active users' is retained as the focal quantity. 

Looking at Figure~1, a clear and important feature is the shrinking growth rate of active users which we model by a relatively flexible ecological-type nonlinear regression,
\begin{equation}\label{eq:eco}
\ln(u)= a-b e^{-c t^d} + \epsilon,
\end{equation}
which saturates at a ``carrying capacity'', $u \rightarrow  e^a$ as $t\rightarrow\infty$, and where the log transform stabilizes the noise level. As in the case of the generalized Metcalfe regression, here there is clear structure in the residuals, as feedback loops develop between the number of active users and price during speculative bubbles. We opt to fit the curve \eqref{eq:eco} by OLS (ordinary least squares) and treat it as a rough estimate: Fitting from 2012-01-01 to 2018-02-26\footnote{Details of the fit: The interval spanned by the natural log of the number of active users was transformed to (0,1) by shifting by 9.483 and dividing by 4.46. The time span was also transformed to (0,1). The nonlinear regression was then fit by OLS, giving parameters and standard deviations a=1.72 (0.14), b= 1.76 (0.15), c=0.79 (0.09), d=0.70 (0.26). Predicted values (transformed back to original scale) for the first day of each year from 2018--2023 in Millions of active users, and percentage standard error are 0.788 (0.05\%), 1.06 (0.06\%), 1.39 (0.07\%), 1.75 (0.07\%), 2.16 (0.074\%), and 2.60 (0.08\%).  Finally, the estimated carrying capacity is $2.76\times10^7$  with standard error of 86\%. }, the annual growth rate is expected to decrease over the next five years from 35\% to 21\%, taking the expected level of active users from 0.79 Million currently to 2.60 Million in 2023 with 5\% and 8\% standard errors, respectively. Comparing with a fit starting earlier, in 2010-10-24\footnote{Doing the same as for the previous fit, but starting from 2010-10-24, gives parameters: $a=2.86~(0.59) b=3.03~(0.61), c= 0.46~(0.11), d= 0.40~(0.02)$, with predicted values for first day of 2018--2023:  0.812, 1.14, 1.54, 2.04, 2.63, and 3.35 (Millions). The predicted growth rates over the next five years are 40\%, 36\%, 32\%, 29\%, and 27\%. And a massive carrying capacity of $9.39 \times 10^{11}$ is predicted with 180\% standard error.}, again a similarly decreasing growth rate is confirmed, but with predictions for 2018 and 2023 respectively being 7\% and 28\% larger than predictions for the first fit. More generally, within the sample, the fitted curves are similar, but, beyond the sample, differences explode such that there are 4 orders of magnitude difference between the predicted carrying capacities. Here, model uncertainty dominates uncertainty of estimated parameters. There is also likely to be some non-stationarity and regime-shifts as the bitcoin network evolves and matures, contributing another level of uncertainty in the long-term extrapolation of our models. Therefore, precise inference based on a single model--- notably omitting any limitation imposed by the physical bitcoin network---is misleading, and long-term predictions effectively meaningless. However, smoothing of past values is not problematic, and short term projections may be reasonable.

\begin{figure}[h!]
\centerline{\includegraphics[width=13cm]{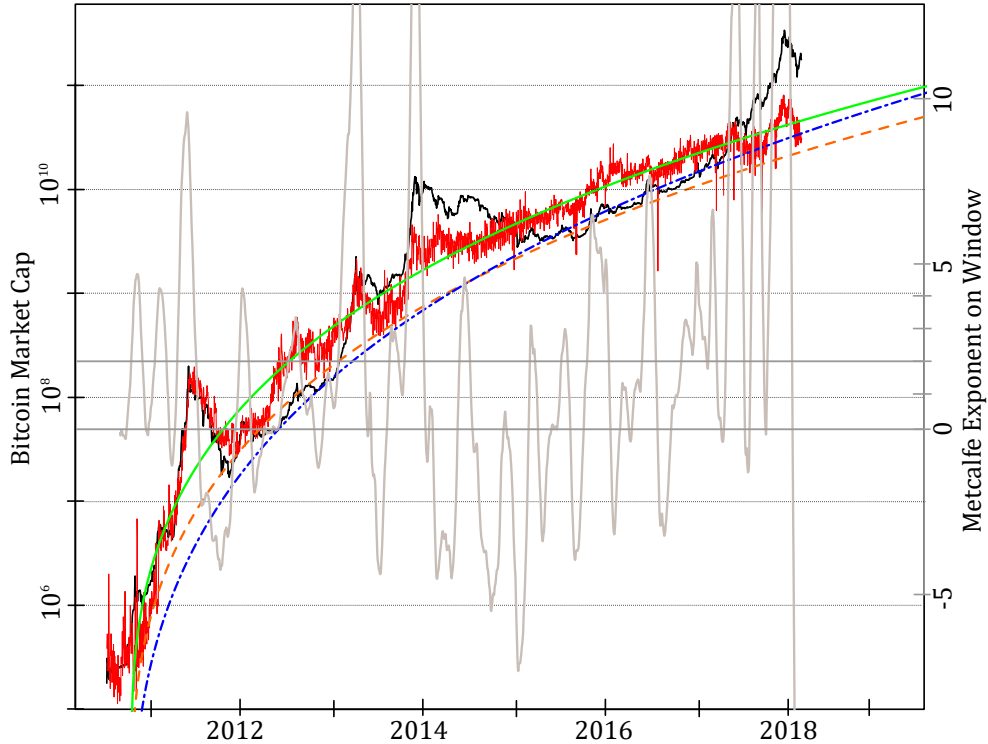}}
\caption{Comparing bitcoin market cap (black line) with predicted market cap based on various generalized Metcalfe regressions of active users. The rough red line is given by plugging the true number of active users into the generalized Metcalfe regression shown in Figure~1, having OLS estimated coefficients $(\alpha,\beta)= (1.51,1.69)$. The remaining lines plug smoothed active users (non-parametric up to 2018 and the nonlinear regression starting in 2012 to project beyond) into the generalized Metcalfe formula with different parameters: The smooth green line for the estimated coefficients (1.51,1.69); the orange dashed line is proposed as a ``support line'', having coefficients (0,1.75) specified by eye; the blue dash-dotted line being a Metcalfe support line with coefficients (-3,2). The grey line, plotted against the right axis, is the exponent of the generalized Metcalfe regression onto smoothed active users on a causal 60 day moving window (i.e., window on the previous 60 days). It is truncated to emphasize fluctuations around the value 2 (solid grey line).}
\label{fig:fig2}
\end{figure}

Given the number of active users, and calibrations of the generalized Metcalfe's law, which maps to market cap, we can now compare the predicted market cap with the true one, as in Figure~2. Also, using smoothed active users, the local endogeneities---where price drives active users---are assumed to be averaged out. The OLS estimated regression, by definition, fits the conditional mean, as is apparent in Figure~2. Therefore, if bitcoin has evolved based on fundamental user growth with transient overvaluations on top, then the OLS estimate will give an estimate in-between and thus above the fundamental value. For this reason, support lines are also given, and---although their parameters are chosen visually---they may give a sounder indication of fundamental value. In any case, the predicted values for the market cap indicate a current over-valuation of at least four times. In particular, the OLS fit with parameters (1.51,1.69), the support line with (0,1.75), and the Metcalfe support line (-3,2) suggest current values around 44, 22, and 33 billion USD, respectively, in contrast to the actual current market cap of 170 billion USD. Further, assuming continued user growth in line with the regression of active users starting in 2012, the end of 2018 Metcalfe predictions for the market cap are 77, 39, and 64 billion USD respectively\footnote{With standard errors already above 10\% induced by estimated parameters, excluding additional prediction uncertainty due to persistent fluctuations of active users about the mean.}, which is still less than half of the current market cap. These results are found to be robust with regards to the chosen fitting window\footnote{Although the parameters vary depending on the fitting window, even allowing for fitting windows starting in 2016, where one obtains a high exponent $\beta$ (above 2.5), an overvaluation of about a factor of two is still indicated.}. 

On this basis alone, the current market looks similar to that of early 2014, which was followed by a year of sideways and downward movement. Some separate fundamental development would need to exist to justify such high valuation, which we are unaware of.

\section{ Bitcoin bubbles: universality of unsustainable growth? }

\subsection{Identification and main properties of the four main bubbles}

Using the generalized Metcalfe regression onto smoothed active users as well as its support lines, one can identify in Figure~2 four main bubbles corresponding to the largest upward deviations of the market cap from this estimated fundamental value. These four bubbles in market cap are highlighted in Figure~3, and detailed in Table 1---in some cases exhibiting a 20 fold increase in less than 6 months! In all cases, the burst of the bubble is attributed to fundamental events, listed below, in particular for the first three bubbles, which corrected rapidly at the time of the clearly relevant news. The fourth and very recent bubble was much longer, and it is plausible that the main news there was really the 20'000 USD value of bitcoin, i.e., it finally collapsed under its own weight\footnote{This large valuation is likely to have attracted ``whales'' to cash a part of their bitcoin portfolios, either to realize their profit or due to operational constraints. For instance, it was revealed on March 2, 2018 that Nobuaki Kobayashi, bankruptcy trustee for Mt. Gox, once the largest bitcoin exchange in the world, has sold off about \$400 million in bitcoin and bitcoin cash since late September 2017 (\url{https://www.zerohedge.com/news/2018-03-07/bitcoins-tokyo-whale-sells-400m-bitcoin-bitcoin-cash}).}. Market participants often lament that crashes are unforeseeable due to the unpredictability of bad news. 

\begin{figure}[h!]
\centerline{\includegraphics[width=13cm]{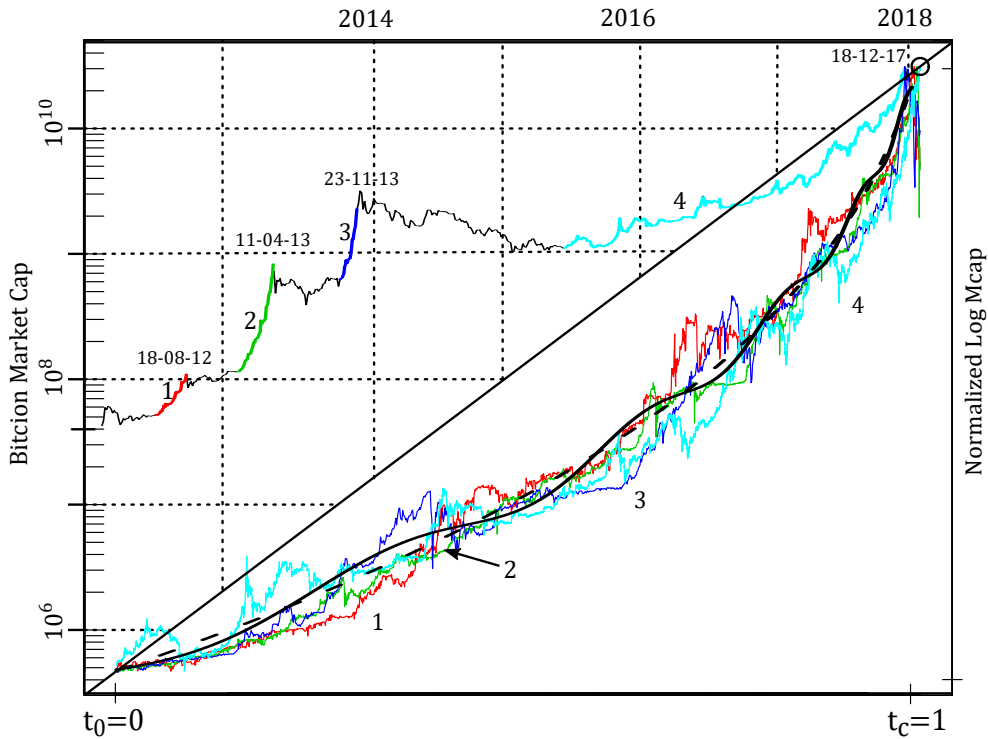}}
\caption{Upper triangle: market cap of bitcoin with four major bubbles indicated by bold colored lines, numbered, and with bursting date given. Lower triangle: The four bubbles scaled to have the same log-height and length, with the same color coding as the upper, and with pure hyperbolic power law and LPPLS models fitted to the average of the four scaled bubbles, given in dashed and solid black, respectively.}
\label{fig:fig3}
\end{figure}

\begin{table}[t!]
\begin{center}
\begin{tabular}{ c | c c c c c c c }
\toprule
Bubble  & Start			&		End		& Days	&M-Cap$_0$&	M-Cap$_1$	&	   Growth &  Mean Return	\\
\midrule	
1 & 	2012-05-25 	&2012-08-18 &  	84 & 	$4.65\times10^7 $	&		$1.45\times10^8$   & 3.1   &  	0.013	\\
2	& 2013-01-03 	&2013-04-11  & 	98  &	$1.39\times10^8 $	&		$2.84\times10^9 $  & 20.4  &	0.031  \\
3	& 2013-10-07 &	2013-11-23  & 	47  &	$1.45\times10^9 $	&		$9.8\times10^9   $  & 6.8  	& 0.042 \\
4	& 2015-06-08  & 2017-12-18  &   924 &	$3.17\times10^9 $	&		$3.27\times10^{11}$  & 103  &	0.005 \\
5	& 2017-03-31  & 2017-12-18  &   155 &	$1.69\times10^{10}$ 	& $3.27\times10^{11}$  &  21  &	0.02	 \\
\bottomrule
 \end{tabular}
\end{center}
\tableCaptionNegativeVSpace\caption{Bubble statistics. Columns: Start, end (time of peak value, prior to correction), duration in days, starting and peak market cap, growth factor (peak divided by start value: M-Cap$_1$ / M-Cap$_0$), and average daily return. The bubbles correspond to the numbering in Figure~3. Bubble 5 corresponds to approximately the last six months of the fourth bubble, and will be used in the next section. The price data for bitcoin is from Bitstamp, in USD, hourly from 2012-01-01 to 2018-01-08; the bitcoin circulating supply comes from blockchain.info. }\label{tab:tab1}
 \end{table}

However, focusing on the news that may have triggered the crash is akin to waiting for ``the final straw'', rather than monitoring the developing  unsustainable load on the poor camel's back. Of particular interest here is that, although the height and length of the bubbles vary considerably, when scaled to have the same log-height and length, a near-universal super-exponential growth is evident, as diagnosed by the overall upward curvature in this linear-logarithmic plot (lower Figure~3). And in this sense, like a sandpile, once the scaled bubble becomes steep enough (angle of repose), it will avalanche, while the precise triggering nudge is essentially irrelevant. 

Below, events thought to trigger crashes/corrections, corresponding to bubbles 1--4 in Table 1 are mentioned\footnote{Events taken from \url{https://99bitcoins.com/price-chart-history/}}:
\begin{enumerate}
\setcounter{enumi}{-1}
 \setlength\itemsep{0.02em}
\item 2011-06-19\footnote{This trigger is for the ``zeroth'' bubble, being before our data window.}: Mt. Gox was hacked, causing the bitcoin price to fall 88\% over the next 3 months. 
\item 2012-08-28: Ponzi fraud of perhaps hundreds of thousands of bitcoin under the name bitcoins Savings \& Trust; charges filed by Securities and Exchange Commission. 
\item 2013-04-10: Major bitcoin exchange, Mt Gox, breaks under high trading volume; price falls more than 50\% over next 2 days. 
\item 2013-12-5: Following strong adoption growth in China, the People's Bank of China bans financial institutions from using bitcoin; bitcoin market cap drops 50\% over the next two weeks. 07-02-2014: operational issues at major exchanges due to distributed denial of service attacks, and two weeks later Mt Gox closes. 
\item 2017-12-28: South Korean regulator threatens to shut down crypto currency exchanges. 
\end{enumerate}

\subsection{Log-periodic finite time singularity model}

Following Sornette and colleagues~\cite{johansen1999predicting,johansen2000crashes,sornette2013clarifications}, as mentioned in the introduction, we consider bubbles to be the result of unsustainable (faster than exponential) growth, achieving an infinite return in finite time (a finite time singularity), forcing a correction / change of regime in the real world. We adopt the LPPLS model, as parameterized in \cite{filimonov2013stable}, for the log market cap, $p_i$ at time $t_i$,
\begin{equation}
\label{eqn:LPPLS}
y_i:=\ln(p_i)=a+(t_c-t_i)^m \Big(b + c \cos\big(w\ln(t_c-t_i)\big) + d \sin\big(w \ln(t_c-t_i)\big) \Big)+\epsilon_i  ,~ t_i
\end{equation}
where $0<m<1$, $\ln(p_c)=a$, and $T_1 \leq t_i<t_c$.  $T_1$ is the starting time, and $t_c$ the stopping or so-called critical time by which the bubble must burst. This model combines two well documented empirical and phenomenological features of bubbles: (1) a transient ``faster-than-exponential'' growth with singularity at $t_c$, modeled by a pure hyperbolic power law (the above equation with $c=d=0$), resulting from positive feedbacks, which is (2) decorated with accelerating periodic volatility fluctuations, embodying spirals of fear and crash expectations.    

The model needs to be fit with data $((y_1,t_1),\dots,(y_n,t_n))$, on a window $(T_1,T_2)$, where $T_1\leq t_1<\dots<t_n \leq T_2<t_c$. The window $(T_1,T_2)$ needs to be specified, with selection of the start of the bubble $T_1$ often being less obvious.
As is typical in time series regression~\cite{greene2000econometric}, the errors $\epsilon_i$ are correlated and may have changing variance (hetero-skedasticity), which if ignored leads to sub-optimal estimates, and confidence intervals that are too small (over-optimistic). In this case, generalized least squares (GLS) provides a conventional solution, which has been used with LPPLS~\cite{gazola2008log,lin2014volatility,vsirca2016jls}  and, if well-specified, has optimal properties. Here, we opt for a simple specification of the error model, being auto-regressive of order 1\footnote{Higher order ARMA models can also be considered, and are seen to leave residuals with little auto-correlation. Given the regression based de-trending, truly long memory in the errors is not expected, and the auto-correlation of residuals is seen to decay clearly faster than a power law. Further, Dickey-Fuller tests tend to reject that the residuals are unit-root, strongly when significant log periodic oscillations are fit.},
\begin{equation} 
\epsilon_i = \phi \epsilon_{i-1} + \eta_i, \quad |\phi|<1,
\end{equation}
to model the rather persistent deviations from the overall trend. We then estimate the LPPLS model by profiling over non-linear parameters $(m,w,t_c,\phi)$, which allows the conditionally linear parameters $(a,b,c,d)$ to be estimated analytically, by GLS, or by OLS if $\phi=0$. Assuming white noise normal errors $\eta_i$, this is maximum likelihood, and allows for profile likelihood confidence intervals of all parameters. 

Here, we focus on $t_c$, the critical time at which the bubble is most likely to burst.
Before taking the Metcalfe fundamental value into account, and to provide a curve to compare with the data in Figure~3, we fit the pure hyperbolic power law (obtained by putting $c=d=0$ in \eqref{eqn:LPPLS}) and the LPPLS model to the average of the four scaled bubbles\footnote{These fits contain future information, in the sense that the end time of each fitted bubble is the time at which the price peaked, which can only be determined after the crash occurred. These fits are thus not for prediction purpose but for assessing the quality of the hyperbolic power law versus LPPLS models.}, with results summarized in Table 2. 
The hyperbolic power law and LPPLS fits provide a similar trend, and the forward-looking predicted critical/bursting time hugs the lower bound of $1.01$ (the true peak being by construction at $1$). 

Perhaps curiously---despite fitting on an average of unsynchronized disparate bubbles with similar overall trajectories---the LPPLS fit is significantly better, based on log-likelihood ($p<10^{-5}$) as it captures some of the persistent fluctuations, and allows for a significantly smaller $\phi$, i.e. a reduction of the memory time $\sim 1/(1-\phi)$ of the residuals by a factor $13$\footnote{This suggests the existence of an intrinsic phase of the log-periodic oscillations with respect to the finite-time rounding
of the mathematical singularity at the market peak before the crash \cite{JohSorCAN98,ZhouSorNP03}.}.

\begin{table}[t!]
\begin{center}
\begin{tabular}{ c c c c c c c c }
\toprule
$a$ & $b$		&		$c$	& $d$	& $\omega$ &	$m$ &	  $t_c$ &  $\phi$	\\
\midrule	
2.00  &		-1.97 &		-0.020 	&	0.013 &		10.79 &		0.23  &		1.03(1.01,1.06)  &		0.87 \\
1.54 &	 		-1.52&		 	=0 			&		=0 	&	  	NA 		&	 	0.31 	&	 	1.02(1.01,1.05) &	 	0.99	 \\
\bottomrule
 \end{tabular}
\end{center}
\tableCaptionNegativeVSpace
\caption{ LPPLS (second row) and pure hyperbolic power law ($c=d=0$) (third row) fits on the average of the four scaled bubbles shown in Figure~3. The sample is taken at 200 equidistant points. The 95\% profile likelihood confidence interval is given for $t_c$.  }\label{tab:tab2}
 \end{table}

\subsection{Bubbles in the Market-to-Metcalfe Ratio}

Given our proposed fundamental value of bitcoin based on the generalized Metcalfe regressions presented above, we define the Market-to-Metcalfe value (MMV) ratio,
\begin{equation}
\text{MMV}_i = \frac{ p_i}{  e^{-3} u_{i}^2},  
\end{equation}
as the actual market cap ($p_i$ at time $t_i$) divided by the market cap predicted by the Metcalfe support level, with parameters ($\alpha_0=-3, \beta_0=2$) in \eqref{eqn:Metcalfslaw}, with smoothed active users ($u_i$) plugged in\footnote{Note that whether the value $\beta=2$ or $\beta=1.75$ are used, the results for this analysis will be effectively identical. }. We sample the value every three hours over the time periods corresponding to bubbles 1--3 and 5 in Table 1.

As shown in Figure~4, bubbles are persistent deviations of the Market-to-Metcalfe value above support level $1$, which are well modelled by the LPPLS model. In particular, the parameters of the hyperbolic power law and LPPLS models fitted on the Market-to-Metcalfe ratio data, for the full bubble lengths, are given in Table 4. For the different bubbles, the key nonlinear parameters fall within similar ranges, and calibration of $t_c$ is accurate. Again, the LPPLS fits dominate the pure hyperbolic power laws, according to likelihood ratios. Further, based on our methodology (see appendix), none of these fits can be rejected on the basis of their residuals.

\begin{figure}[H]
\centerline{\includegraphics[width=18cm]{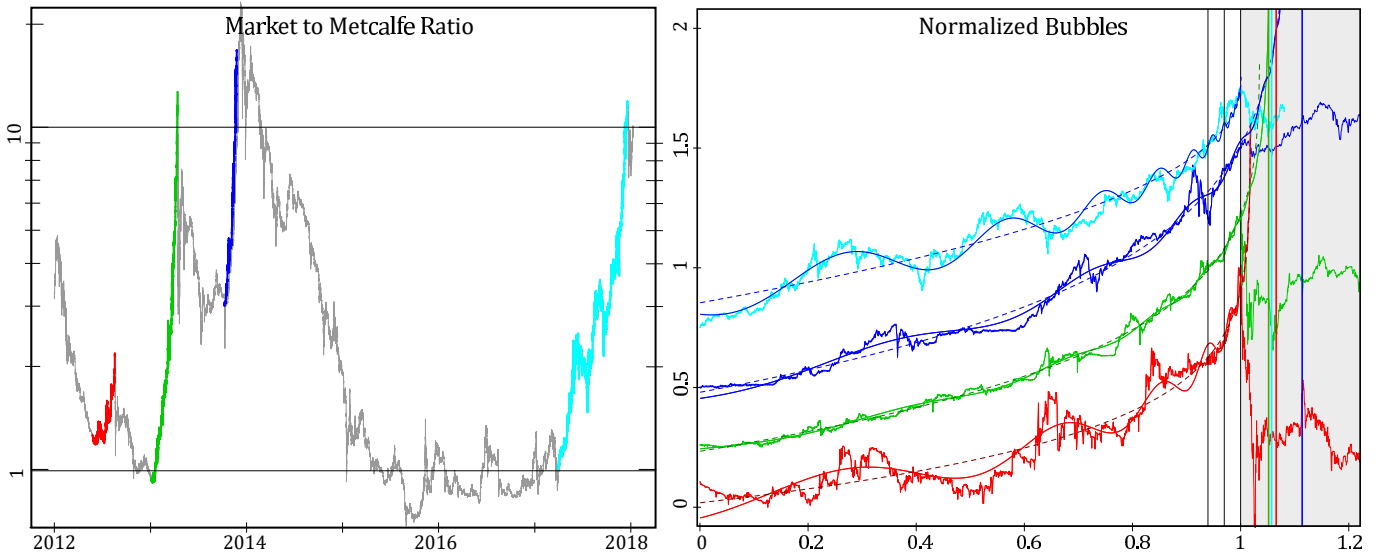}}
\caption{Left panel: Market-to-Metcalfe value ratio (MMV) over time. The apparent bubbles, which radically depart from the fundamental level 1, are colored and given in Table 1 as bubbles 1--3 and 5. Right panel: for the same four bubbles, the MMV ratios are shown in log-scale
as a function of linear rescaled time, with 0.25 vertical offset for visibility.  The hyperbolic power law and LPPLS fits on the 4 full bubbles are shown. Values of the MMV ratio after the bubble peak are shown on the grey background, where the colored vertical lines indicate the upper limit 
for $t_c$ of the 95\% profile likelihood confidence interval for each of the four bubbles. The three thin vertical black lines gives the rightmost edge of the 95, 97.5, and 100\% data windows on which fits were done, with parameters summarized in Table 3 and Appendix Table 5. }
\label{fig:fig4}
\end{figure}

The ex-ante predictive aspect is important as, in addition to verifying the LPPLS bubble in hindsight, one would like to have a sound advance warning of the bubble's existence and a reasonable confidence interval for its bursting time. Here, we provide a simple indication of this potential with two additional sets of fits: fitting with bubble data up to 95\% and 97.5\% of the bubble length. The overall parameter estimates (see appendix Table 6) are similar to the 100\% window, in Table 4, with key nonlinear parameters typically in ranges $0.1<m<0.5$, and $7<w<11$. Focusing on the critical bursting time, in Table 3, the estimated $t_c$ and 95\% confidence intervals are given, showing quite stable advance-warning. That is, point estimates and confidence intervals are consistent with the true bursting time, noting that $t_c$ is in theory both the most probable and latest time for the burst of the bubble \cite{johansen1999predicting,johansen2000crashes,sornette2013clarifications}, as the market is increasingly susceptible as it approaches $t_c$, and can therefore be toppled by bad news.

\begin{table}[h]
\begin{center}
\begin{tabular}{ c | c c c c c c c c }
\toprule
Fit &a & b		&		c	& d	& w &	m &	  $t_c$ &  $\phi$	\\
\midrule	
 1 & 2.74 &-2.72& 	-0.051& -0.044 & 	8.37  &	0.10  &	1.02 (1.01,1.09)& 	0.92 \\
 2  &3.74 &-3.73	& -0.005&  0.012  	&10.80  & 	0.10&  	1.05 (1.03,1.06) &	0.90\\
 3  &4.56 &-4.53	& -0.031& -0.013  &	8.97& 	 0.10  &	1.09 (1.05,1.12)& 	0.92\\
 5  &1.09 &-0.96 &	-0.071&  0.053 	&12.00  &	0.38 & 	1.01 (1.01,1.07)& 	0.98\\
 \midrule	
1a  &2.71 &-2.68  &	=0&	=0	&NA&	0.10  &	1.02 (1.01,1.04) &	0.98 \\
 2a  &2.13 &-2.13  &	=0&	=0&	NA&	0.18  &	1.04 (1.02,1.04) &	0.99 \\
 3a &4.61 &-4.59  &	=0&	=0&	NA&	0.10  &	1.09 (1.05,1.23) &	0.97 \\
 5a  &1.046& -0.94& 	=0&	=0&	NA&	0.43 & 	1.00 (1.01,1.20) &	0.99 \\
\bottomrule
 \end{tabular}
\end{center}
\tableCaptionNegativeVSpace
\caption{Estimated parameters of the LPPLS and hyperbolic power law models on the Market-to-Metcalfe value ratios for the four bubbles, indicated by the fit number. The suffix `a' corresponds to the hyperbolic power law fits of the Market-to-Metcalfe value ratios for these four bubbles.
The 95\% profile likelihood confidence interval for $t_c$ is given. The likelihood ratio test of the LPPLS versus the hyperbolic power law (null) gives p-values of $0.01$, $10^{-5}$, $0.02$, and $0.07$, for these four bubbles.  A lower bound for m of 0.1 was enforced. }\label{tab:tab3}
 \end{table}

\begin{table}[t!]
\begin{center}
\begin{tabular}{ c | c c c  }
\toprule
Fit & 0.95	 &		0.975 &			1	\\		
\midrule	
 1		& 0.99 (0.98,1.08) 	&	 1.01 (0.99,1.08) 	&	 1.02 (1.01,1.09)   \\		
 2		&0.99 (0.98,1.0) 		& 1.07 (1.05,1.07) 	 	&1.05 (1.03,1.06) \\		
 3		& 1.02 (1.01,1.02) 	& 1.07 (1.04,1.08)	 	&1.09 (1.05,1.12) \\		
 5	 	&0.97 (0.97,0.98) 		& 0.98 (1.01,1.06) 	 	&1.00 (1.01,1.07) \\		
1a	& 0.99 (0.97,1.4)  	&  1.01 (0.98,1.32) 	 	&1.02 (1.01,1.04)\\		
2a	& 1.00 (0.99,1.04) 	& 1.06 (1.04,1.11)  		&1.04 (1.02,1.04)\\		
3a 	&1.08 (1.0,1.4)    	 	& 1.08 (1.01,1.25) 		&1.09 (1.05,1.23)\\		
5a	& 0.95 (0.95,1.4)  	& 0.98 (0.98,1.4) 		&1.00 (1.01,1.20)	\\		
\bottomrule
 \end{tabular}
\end{center}
\tableCaptionNegativeVSpace
\caption{ Estimated critical time and 95\% confidence interval, for LPPLS and hyperbolic power law fits of the Market-to-Metcalfe value 
ratios of the four bubbles, indicated by the fit number and suffixed with a, as defined in Table 3.
The three columns are for fits on data up to $T_2$, being 95, 97.5, and 100\% of the bubble length, as indicated by bubbles 1--3 and 5 in Table 1. }\label{tab:tab4}
\end{table}

\pagebreak
\newpage

\section{Discussion}
 
In this paper, we have combined a generalized Metcalfe's law, providing a fundamental value based on network characteristics, with the Log-periodic Power law Singularity (LPPLS) model, to develop a rich diagnostic of bubbles and their crashes that have punctuated the cryptocurrency's history. In doing so, we were able to diagnose four distinct bubbles, being periods of high overvaluation and LPPLS-like trajectories, which were followed by crashes or strong corrections. Although the height and length of the bubbles vary substantially, we showed that, when scaled to the same log-height and length, a near-universal super-exponential growth is documented. This is in radical contrast to the view that crypto-markets follow a random walk and are essentially unpredictable.

Further, in addition to being able to identify bubbles in hindsight, given the consistent LPPLS bubble characteristics and demonstrated advance warning potential, the LPPLS can be used to provide ex-ante predictions. For instance, a reasonable confidence interval for the endogeneous critical time indicates a high hazard for correction in that neighborhood, as any minor event could topple the unstable market. Of course, massive exogeneous shocks, although rare, could occur at any time, and the LPPLS model can provide no warning there.

Focusing on the outlook for bitcoin, the active user data indicates a shrinking growth rate, which a range of parameterizations of our
generalized Metcalfe's law translates into slowing growth in market capitalization. Further, our Metcalfe-based analysis indicates current support levels for the bitcoin market in the range of 22--44 billion USD, at least four times less than the current level. On this basis alone, the current market resembles that of early 2014, which was followed by a year of sideways and downward movement. Given the high correlation of cryptocurrencies, the short-term movements of other cryptocurrencies are likely to be affected by corrections in bitcoin (and vice-versa), regardless of their own relative valuations.

\pagebreak

\bibliographystyle{unsrt}
\bibliography{bitcoinBubble3}

\pagebreak

\section{Appendix}

\begin{table}[h!]
\begin{center}
\begin{tabular}{ c | c c c c c c c c }
97.5\% &&&&&&&& \\
Fit &a & b		&		c	& d	& w &	m &	  $t_c$ &  $\phi$	\\
\midrule	
 1&  2.72		& -2.72		&-0.035		& -0.062 	& 7.76  		&	0.10  		&	1.01 (0.99,1.08) 		&0.92 \\
 2 & 3.88		& -3.87		& 	-0.004 	& 0.013		& 11.39 		& 0.10 		&	1.07 (1.05,1.07) 		&	0.90 \\
 3&  4.32 	&-4.30 		&	-0.035 	&-0.012 		& 8.37  		&0.10  		&1.07 (1.04,1.08) 		&	0.92 \\
 5 & 0.91		& -0.79		&	 -0.062 	& 0.074		& 11.34 		& 	0.48  		&	0.98 (1.01,1.06)		& 	0.98 \\
 1a & 2.62	& -2.60		&=0				&	=0			&	NA			&	 0.10 		&	1.01 (0.98,1.32)			& 	0.96   \\
 2a & 3.88	& -3.87  		&	=0			&	=0			&	NA			&	0.10  		&	1.06 (1.04,1.11) 		&	0.98   \\
 3a & 4.58	& -4.56  		&	=0			&=0				&NA			&0.10  		&	1.08 (1.01,1.25) 		&	0.97  	 \\
 5a &0.86	&-0.77  		&	=0			&	=0			&	NA			&0.58 			& 0.98 (0.98,1.4) 		&	0.99  	  \\
 \midrule	
95\% &&&&&&&& \\
Fit &a & b		&		c	& d	& w &	m &	  $t_c$ &  $\phi$	\\
\midrule	
 1&  2.53		& -2.52 	&	-0.059	& -0.040		&  7.76  		&	0.10 	& 	0.99 (0.98,1.08)	 	& 0.92 \\
 2 & 1.42		& -1.44		&  0.013	&  0.007		& 10.79  		&	0.26  	&	0.99 (0.98,1.0)  		& 0.90 \\
 3 & 2.96		& -2.95 	&	-0.004	&  0.043		& 10.18  		&	0.13 	&	1.02 (1.01,1.02) 	&	0.93 \\
 5 & 0.81		& -0.71		&-0.054	&  0.085		& 11.40 	&	0.57  	&	0.97 (0.97,0.98)		 & 0.96 \\
 1a&2.47	& -2.45		&	=0		&=0				&NA			&  0.10  &	0.99 (0.97,1.4) 	& 0.96  \\
 2a&  1.54	& -1.55 	&=0			&	=0			&NA			&	0.25  &	1.00 (0.99,1.04) 	& 0.98 \\
 3a&  4.60	& -4.58  		&=0			&=0				&	NA			&0.10 		& 	1.08 (1.0,1.4)	 		& 0.96  \\
 5a&  0.75	&-0.68  	&	=0		&=0				&	NA			&	0.70  &	0.95 (0.95,1.4) 	& 0.99  \\
\bottomrule
 \end{tabular}
\end{center}
\tableCaptionNegativeVSpace
\caption{The same as Table 3, but fits up to 95\%, and 97.5\% of the bubble length, rather than 100\%. The likelihood ratio test p-values of bubbles 1--3 and 5, with the pure hyperbolic power law fit as the null, for the first sub-table are 0.05, 0. 0007, 0.01, and 0.02; and for the second sub-table, 0.05, $<10^{-6}$, 0.0004, and 0.003.  }\label{tab:tab6}
 \end{table}

 \subsection{LPPLS algorithm}
A rough algorithm for fitting LPPL is given, and illustrated with a data example in Figure~5. Assumed are existence of a smooth trend in a window before the finite time singularity at $t_c$, and that a stationary time series model exists for the---often persistent---errors around that trend. It allows for selection of a best window, giving the bubble starting time, $T_1$, by a hypothesis test, and confidence intervals for the critical time $t_c$, which are more realistic than if assuming iid errors.

\begin{figure}[h!]
\centerline{\includegraphics[width=12cm]{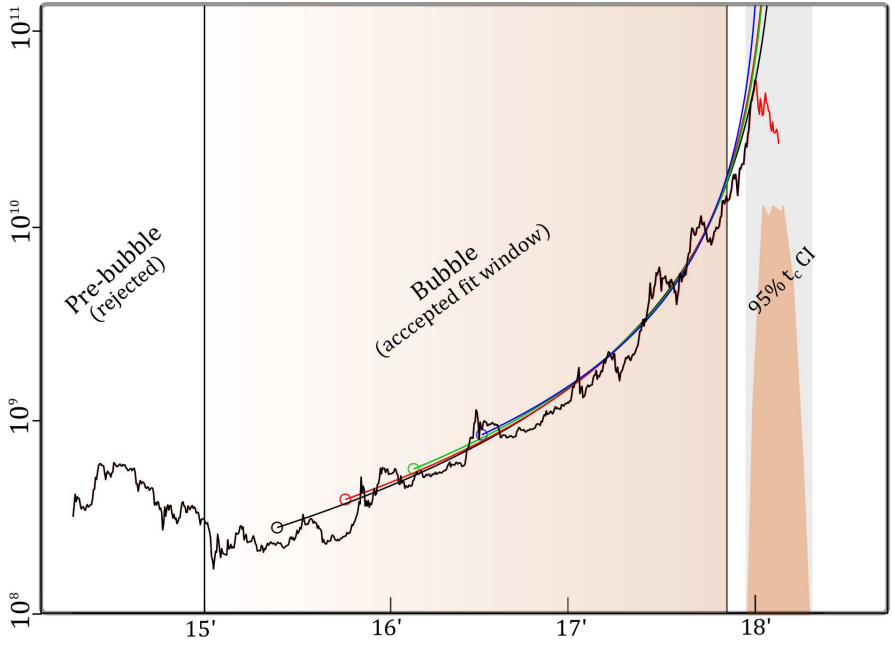}}
\caption{ 2015--2018 bitcoin market cap bubble to serve as an illustration for the algorithm. Plotted are four accepted pure power law regressions, with upper limit of the fitting window $T_2$ placed 2 months prior to the turning point, and with four different values of the bubble starting time, $T_1$. The orange mode is the average of the four profile likelihoods for $t_c$ for the four fits shown up to the 95\% level, bounded by the light grey bar, giving the 95\% interval.}
\label{fig:fig5}
\end{figure}

\subsubsection*{Steps}
\begin{enumerate}
	\item Identify initial error model: Take broad window $(T_0,T_2)$ thought to contain the bubble, $ T_0<T_1<T_2<t_c$. Fit the log market cap with a flexible non-parametric curve to obtain an estimate of the trend. We use the R \texttt{loess} library and Akaike Information Criterion (AIC) to select degrees of freedom.  Then fit the error-model, here an AR(1) time series, onto the de-trended data, giving an initial estimate for the GLS LPPLS estimation.\\
	\item Characterize error variance: Bootstrap the residuals from step 1 and feed them through the fitted AR(1) to simulate errors, allowing for the distribution of the residual standard error on different window sizes to be approximated by Monte Carlo. Due to the autocorrelated errors, a chi-square distribution will not be valid. \\
	\item Fit LPPLS function by profile-likelihood with GLS: Given a fitting window $(T_1,T_2)$, take a fine grid of nonlinear parameters $(m,w, t_c)$,  and for each point do a GLS fit with, in this case AR(1) errors, initialized from step 1. A maximum likelihood implementation of this is given in R:gls, and detailed in Ch. 5 of~\cite{pinheiro2000linear}, which internally profiles over the AR(1) parameter. An iterative re-weighting to estimate the AR(1) parameter is also an option. Then, take the fit with the highest log-likelihood of all fits. One may use whatever numerical optimization algorithm, but the grid search easily allows for profile likelihoods to be computed. \\
	\item Perform the fit on many windows and choose the best: Here, varying bubble start $T_1$, where $T_0<T_1<T_2$, repeat step 3. For each fit, having sample size n, take the residual error, $RSS/(n-p)$, where p is the degrees of freedom of the LPPLS (take $p=7$ as an upper bound), and RSS is the residual sum of squares. Then compare this value with the distribution of residual errors generated from step 2, possibly bootstrapping only from the fitted window $(T_1,T_2)$ rather than the overall window $(T_0,T_2)$  which may having unbalanced variance. Then for a single fit, take the fit on the largest window that is not rejected. For robustness, one may also wish to consider multiple non-rejected fits. The same approach can be used to select $T_2$, which although often visually obvious, can then be identified in an objective automated way.
\end{enumerate}

\end{document}